 \documentclass[final,5p,times,twocolumn,authoryear]{elsarticle}

\usepackage{amssymb}
\usepackage[hidelinks]{hyperref}

\def\lsim{\lower.5ex\hbox{$\; \buildrel < \over \sim \;$}}
\def\gsim{\lower.5ex\hbox{$\; \buildrel > \over \sim \;$}}

\journal{New Astronomy Reviews}

\begin{document}

\begin{frontmatter}



\title{\emph{INTEGRAL} serendipitous observations of solar and terrestrial X-rays and gamma rays}
%%\title{Solar and terrestrial physics with the \emph{INTEGRAL} space mission} %% suggested by VT
%%\title{Serendipitous Science: Aurorae, Solar Flares, Radiation Belts} %% originally proposed


\author[MT]{Marc T\"urler\corref{scnat}}\ead{marc.turler@unige.ch}
\author[VT]{Vincent Tatischeff}\ead{vincent.tatischeff@csnsm.in2p3.fr}
\author[VB]{Volker Beckmann}
\author[EC1,EC2]{Eugene Churazov}

%% \address{}

\address[MT]{Department of Astronomy of the University of Geneva, ch. d'Ecogia 16, 1290 Versoix, Switzerland}
\cortext[scnat]{Present address: Swiss Academy of Sciences (SCNAT), Haus der Akademien, Laupenstrasse 7, 3008 Bern, Switzerland}
\address[VT]{Universit\'e Paris-Saclay, CNRS/IN2P3, IJCLab, F--91405, Orsay, France}
\address[VB]{Minist\`ere de l'Enseignement Sup\'erieur, de la Recherche et de l'Innovation, 1 rue Descartes, 75005 Paris, France}
\address[EC1]{Max-Planck-Institut f\"{u}r Astrophysik (MPA), Karl-Schwarzschild-Strasse 1, Garching 85741, Germany}
\address[EC2]{Space Research Institute (IKI), Profsoyuznaya 84/32, Moscow 117997, Russia}

\begin{abstract}
ESA's \emph{INTEGRAL} space mission has achieved unique results for solar and terrestrial physics, although spacecraft operations nominally excluded the possibility to point at the Sun or the Earth. The Earth avoidance was, however, exceptionally relaxed for special occultation observations of the Cosmic X-ray Background (CXB), which on some occasions allowed the detection of strong X-ray auroral emission. In addition, the most intense solar flares can be bright enough to be detectable from outside the field of view of the main instruments. This article presents for the first time the auroral observations by \emph{INTEGRAL} and reviews earlier studies of the most intense solar flares. We end by briefly summarising the studies of the Earth's radiation belts, which can be considered as another topic of serendipitous science with \emph{INTEGRAL}.
\end{abstract}




\end{frontmatter}


\section{Introduction}
\label{sec:introduction}

Opening a new window on the sky always leads to surprises. There are the foreseeable discoveries, like the detection of new sources or possibly of a new class of sources, but also the really unexpected. For ESA's \emph{INTEGRAL} mission \citep{WCD03}, this serendipitous science includes solar and terrestrial physics, because observational constraints excluded pointing towards the Sun or the Earth in nominal operation. An exception was made for the study of the Cosmic X-ray Background (CXB) with the use of the Earth as a shield to occult this diffuse emission in part of the \emph{INTEGRAL} field of view \citep{CSR07,TCC10}. In practice, this turned out to be quite challenging due to the fact that the Earth is not dark at hard X-rays. There are even three distinct emission components contributing to the overall X-ray emission: the CXB reflection, the cosmic-ray induced atmospheric emission, and X-ray aurorae. The latter could be neglected in the first series of Earth-occultation observations in 2006 during solar minimum. In subsequent campaigns at solar maximum, there have been two particular observations with a very strong variable emission up to hard X-rays, which we present here for the first time in Sect.~\ref{sec:aurorae}.

Sect.~\ref{sec:solar_flares}  is devoted to solar flares, which are huge explosions at the surface of the Sun. Although they are very intense X-ray and gamma-ray emitters, solar flares were disregarded or thought as impossible to observe, because the \emph{INTEGRAL} spacecraft design imposes to always point at more than 50 degrees away from the Sun. Despite this, both the IBIS and the SPI instruments can detect solar gamma rays from the most intense flares entering the instruments from the side or the back (see Fig.~\ref{fig:solar1}). We review here the results obtained for a few of these extremely energetic flares, which could be detected up to photon energies of about 10~MeV.

As part of serendipitous science, one should also mention the study of the radiation belts, which we briefly address in Sect.~\ref{sec:radiation_belts}. The highly elliptical orbit of the \emph{INTEGRAL} satellite makes it cross the radiation belts at every perigee passage. The \emph{INTEGRAL} Radiation Environment Monitor (IREM) aboard the spacecraft is a particle counter \citep{HBE03}. The progressive change of the orbit of \emph{INTEGRAL} over the years allowed to study the shape and properties of the radiation belts.

\section{Aurorae}
\label{sec:aurorae}

\subsection{Observational context}
\label{sec:observational_context}
After a successful campaign of CXB observations via occultation by the Earth in January-February 2006 \citep{CSR07,TCC10}, a second set of \emph{INTEGRAL} observations was proposed in 2010 and subsequent years. The motivation for additional Earth observations (EOs) was not only to increase the statistics, but also to benefit from a higher solar activity compared to 2006. Indeed, when the solar activity increases, the heliosphere expands and would reduce the incoming flux of low-energy cosmic rays. This would, in turn, lead to a reduced brightness of the cosmic-ray induced atmospheric emission, which strongly affects the determination of the CXB at energies above $\sim$70\,keV \citep{SCS07}.  A first series of five new EOs was performed in 2012 \citep{HLS11}. Unfortunately, their analysis showed several issues blending the signal of the CXB occultation in the detector counts \citep{TPP13}. The problems have been identified to be of three types: 1) a radioactive spacecraft decay after radiation belts exit, 2) a variable instrumental background from the solar wind, and 3) auroral X-ray emission. Whereas the two latter issues are a consequence of higher solar activity around the solar maximum, the first one is related to the progressive change of the \emph{INTEGRAL} orbit, which is now crossing the proton belt. This issue was solved later on by performing the EOs at the end of the revolution (pre-perigee) instead of at its start (post-perigee). In total, four pre-perigee Earth observations were performed in July 2013, December 2013, November 2015 and May 2016. The strongest X-ray aurorae have been observed during EO 2.5 on 20 November 2012 and EO 4.1 on 10 November 2015. The results were never published before, although they have been presented orally at various occasions and a stunning movie of the aurorae in hard X-rays has been released on-line by ESA on 26 January 2016\footnote{\href{https://sci.esa.int/web/integral/-/57258-integral-x-ray-view-of-earth-aurora}{sci.esa.int/web/integral/-/57258-integral-x-ray-view-of-earth-aurora}}.

\subsection{Data analysis}
\label{sec:data_analysis}

The Offline Scientific Analysis (OSA) software package of the \emph{INTEGRAL} mission was built to analyse point sources and is therefore not suited for the study of diffuse emission or of extended sources, which require specific analysis methods. We only used the OSA 10.2 software provided by the \emph{INTEGRAL} Science Data Centre (ISDC, \citet{CWB03}) to extract light-curves of detector counts in different energy bins. For the IBIS/ISGRI instrument \citep{LLL03}, OSA is run from the level COR to DEAD and the light-curves are then produced with \texttt{ii\_light} in time bins of 60 seconds and in 12 energy bins: 20--24, 24--28, 28--32, 32--38, 38--44, 44--52, 52--60, 60--72, 72--84, 84--100, 100--120, and 120--160\,keV. For the JEM-X instrument \citep{LBW03}, OSA is run from the level COR to BIN\_T --- optionally including `slew' science windows --- with a time step of 60 seconds in six energy bins: 3--5, 5--8, 8--12, 12--18, 18--24, 24--30 keV.

\begin{figure}[t]
\begin{center}
\includegraphics[width=1.0\linewidth]{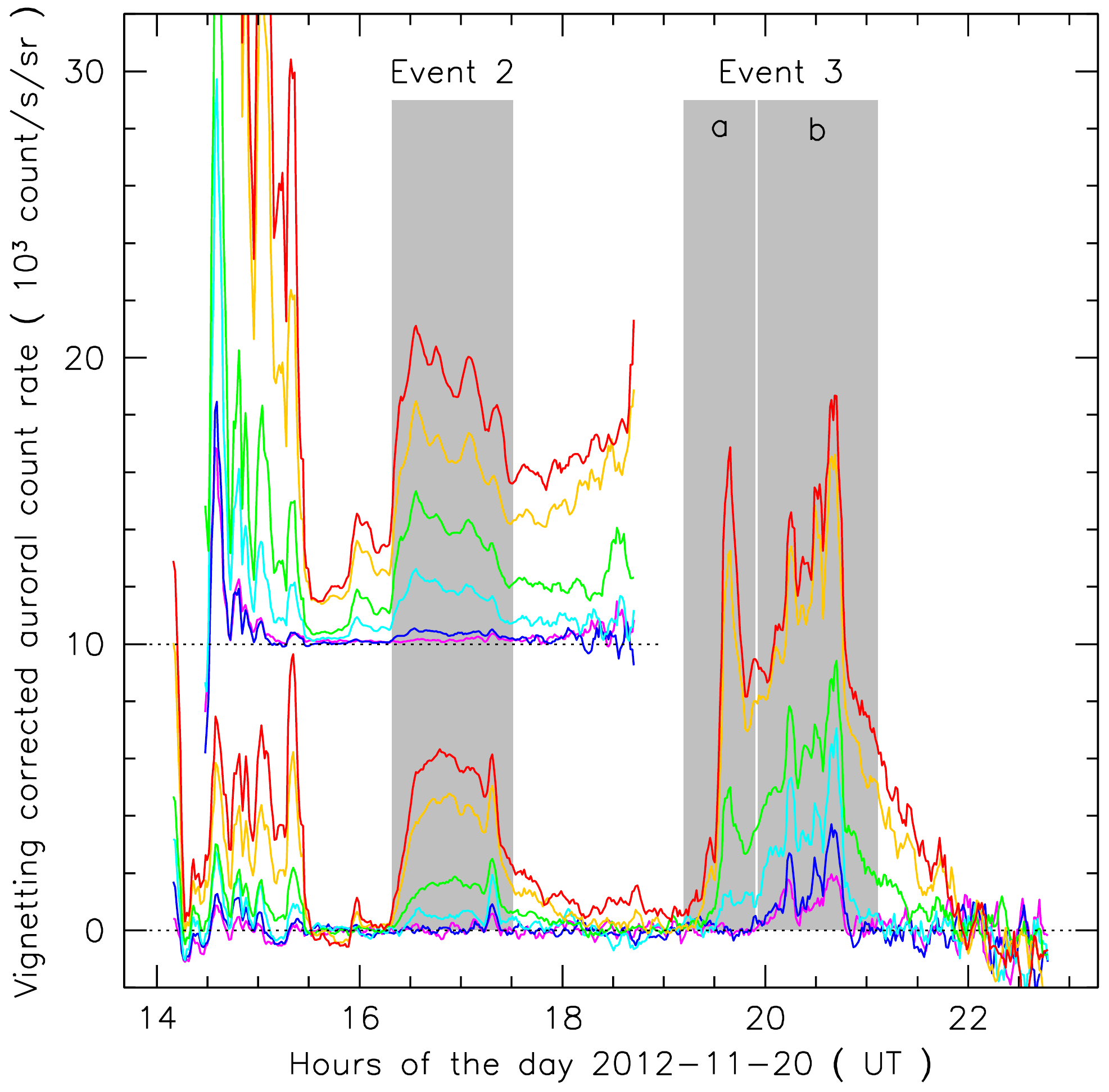}
\end{center}
\caption 
{ \label{fig:lightcurves2012}
Auroral count-rate light-curves in IBIS/ISGRI and JEM-X on 20 November 2012. The JEM-X light-curves in the six energy bins are shown from the lowest to the highest energies in the rainbow colour sequence: red, yellow, green, cyan, blue, and purple. They are smoothed with a time window of 5 minutes and have the origin (dotted-line) vertically displaced for the sake of clarity. They end before 19:00 UT, because the Earth is more rapidly drifting outside of its smaller field of view. The six ISGRI light-curves correspond to pairs of adjacent energy bins (20--28, 28--38, 38--52, 52--72, 72--100, 100--160 keV). Alike for JEM-X, they are smoothed on a timescale of 5 minutes and are drawn with the same rainbow colour sequence. The grey-shaded areas correspond to the time intervals used to extract the distinct auroral spectra (see  Fig.~\ref{fig:spectra2012}).
}
\end{figure}

\begin{figure}[t]
\begin{center}
\includegraphics[width=1.0\linewidth]{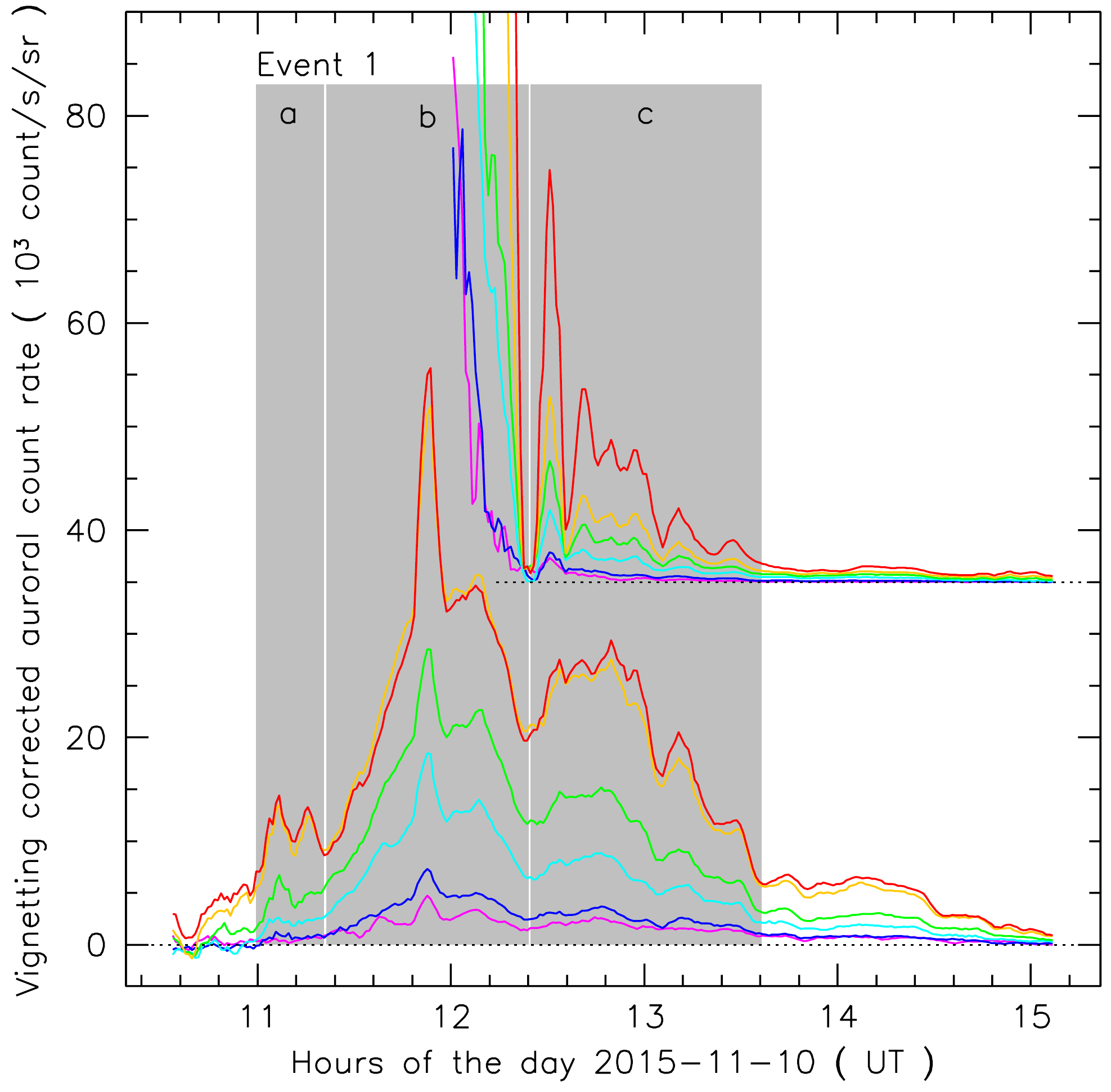}
\end{center}
\caption 
{ \label{fig:lightcurves2015}
Auroral count-rate light-curves in IBIS/ISGRI and JEM-X on 10 November 2015. Energy bins and corresponding colours are as in Fig.~\ref{fig:lightcurves2012}. Here again, the light-curves are smoothed on a timescale of 5 minutes. The JEM-X light-curves are here divided by a factor of 3 and again vertically displaced for clarity. Their very high count rate is probably not reliable before 12.4 UT, because of the emission region just entering the field of view, such that any reflection would be over-amplified by the very strong vignetting correction. The grey-shaded areas correspond to the time intervals used to extract the distinct auroral spectra (see  Fig.~\ref{fig:spectra2015}).
}
\end{figure}

Strong auroral emission can easily be identified in these full-detector count rates as episodes of strong variability on timescales of minutes to hours, which fully dominate the emission in the lower energy bins of both instruments, due to the soft spectrum of the auroral emission. The aurorae are so bright during EO 2.5 and EO 4.1 that one can treat the much weaker contributions from point sources, the CXB, and the Earth as indistinct from the instrumental background. In the post-perigee EO 2.5 this underlying component has primarily the shape of an exponential decay due to the radioactive deactivation of the spacecraft. We therefore identify in the light curves time periods when the auroral emission seems to have returned to full quiescence and fit these data with an exponential curve of the form $a+b\cdot\exp(-(t-t_0)/c)$, where $a$, $b$, and $c$ are free parameters of the fit. After subtracting this background emission, we get full detector count rates, which need to be further corrected for vignetting. The vignetting is time-dependent for the auroral emission, because the Earth is moving, changing size and orientation in the field of view of the instrument. An accurate modelling of the different vignetting components and of the emission region on the surface of the Earth is rather complex. For IBIS/ISGRI, we reuse the same methodology as in \citet{TCC10} and simply model the auroral emission on the visible surface of the Earth by a gaussian ring centred at a geomagnetic latitude of 65$^{\circ}$ and with a longitudinal width of $\sigma = 5^{\circ}$. The normalisation is set to 1\,count\,s$^{-1}$\,sr$^{-1}$ on average over the whole Earth disk when looking directly onto the geomagnetic pole. The same is done for JEM-X, where we simply apply the vignetting correction map used by the OSA software. The resulting, vignetting-corrected light-curves are shown in Figs.~\ref{fig:lightcurves2012} and~\ref{fig:lightcurves2015}.

We obtain auroral spectra by averaging the count rates in specific time intervals, which are visually shown by the grey areas in Figs.~\ref{fig:lightcurves2012} and~\ref{fig:lightcurves2015}. Statistical errors are averaged in quadrature and we further add a systematic uncertainty of 5\,\%, which is justified given the complex, non-standard method we use. The count rate spectra are then written into FITS files for spectral fitting in XSPEC \citep{A96}. When we can combine simultaneous JEM-X and ISGRI data, we obtain a best fit of the curved spectrum with two \emph{bremsstrahlung} components. A single bremsstrahlung model or a cut-off power-law are not good alternatives. The two first energy bins of ISGRI, below 28\,keV, were found to be incompatible with the overall spectral shape and were always ignored in the spectral analysis. This is a well-known limitation of the reliable spectral range of ISGRI. We note that the normalisation of the standard bremsstrahlung model assumes solar abundance of the gas, such that its actual value would differ for the very different composition of the terrestrial atmosphere.

\begin{figure}[t]
\begin{center}
\includegraphics[width=1.0\linewidth]{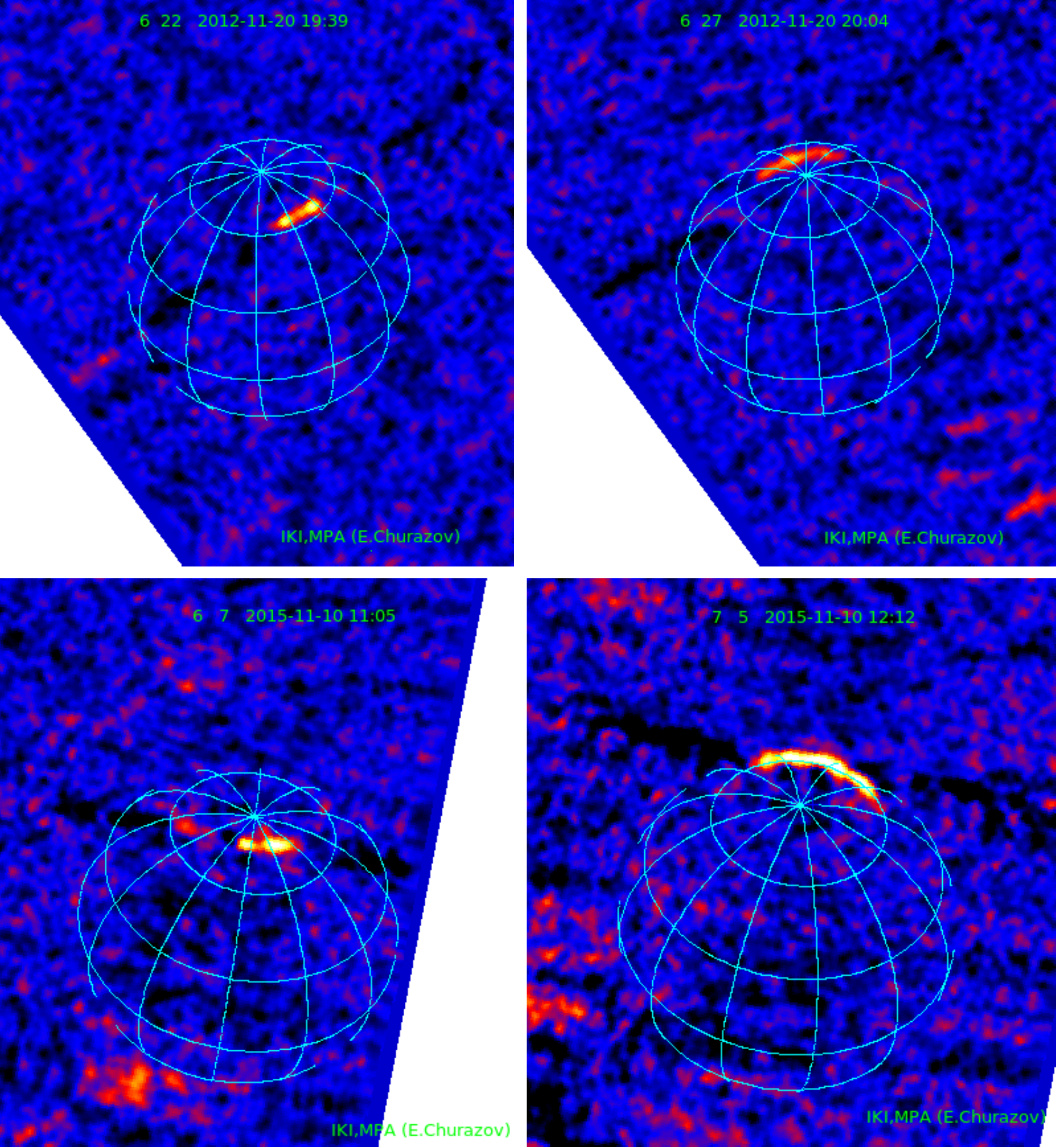}
\end{center}
\caption 
{ \label{fig:aurora_images}
Four images of aurorae obtained with the IBIS/ISGRI instrument in the 17--60 keV energy band in short time intervals. The position of the Earth is indicated by its overlaid coordinate grid. The two upper images are for the aurora of 20 November 2012 at times of 19:39 and 20:04 UT and the two lower ones for the aurora of 10 November 2015 at times of 11:05 and 12:12 UT. The two images on the left-hand side correspond to times during the onset of the event, whereas the right-hand side images are taken later, near the maximum of the event. Interestingly, the onset comes from a more compact region, before the emission quickly moves to a bright arc on the opposite side of the Earth.
}
\end{figure} 

In additional to light-curves and spectra, it was also possible to extract images of the aurorae with the ISGRI instrument. Because the Earth moved through the field of view of the \emph{INTEGRAL} instruments, while the spacecraft moved on its elongated orbit and pointed on a fixed sky position, both the Earth location and its apparent size on the detector are changing rapidly. To cope with these changes, the data of IBIS/ISGRI were chopped into intervals of 500\,s. For each interval, an image in the broad 17--60 keV band has been reconstructed and subsequently convolved with the effective Point Spread Function (PSF) of the telescope, which size reflects the angular size of the IBIS mask elements $\sim 12'$.

A selection of these images corresponding to time periods of particular interest are shown in Fig.~\ref{fig:aurora_images}. The grid of terrestrial latitudes and longitudes has been generated separately for each time interval based on the spacecraft attitude and orbit data and imported into the commonly used 'ds9' visualisation tool as a so-called 'region' file to be overlaid onto the image. This allows to verify that the bright spots and arcs are indeed located in the auroral region at a latitude of $\sim 65^{\circ}$ North. The dark bands running diagonally through the image are artefacts induced by the bright emission of the aurora. They appear due to the periodic pattern of the IBIS coded mask and cannot be easily removed, because the standard method to do so, as used in the IBIS data analysis \citep{GDF03}, does only work for point sources and not for such extended emission regions.

It is worth noting that we also attempted to perform an analysis of the data obtained with the SPI instrument \citep{VRS03}. Preliminary results are consistent with the light-curve profile and the overall spectral shape of the auroral event of EO 4.1 presented below. As the SPI data do not have the potential to add much information to the present JEM-X and IBIS analysis, we did not pursue the analysis up to the level needed for publication.

%%% AURORA 2012 %%%

\begin{figure}[t]
\begin{center}
\includegraphics[width=1.0\linewidth]{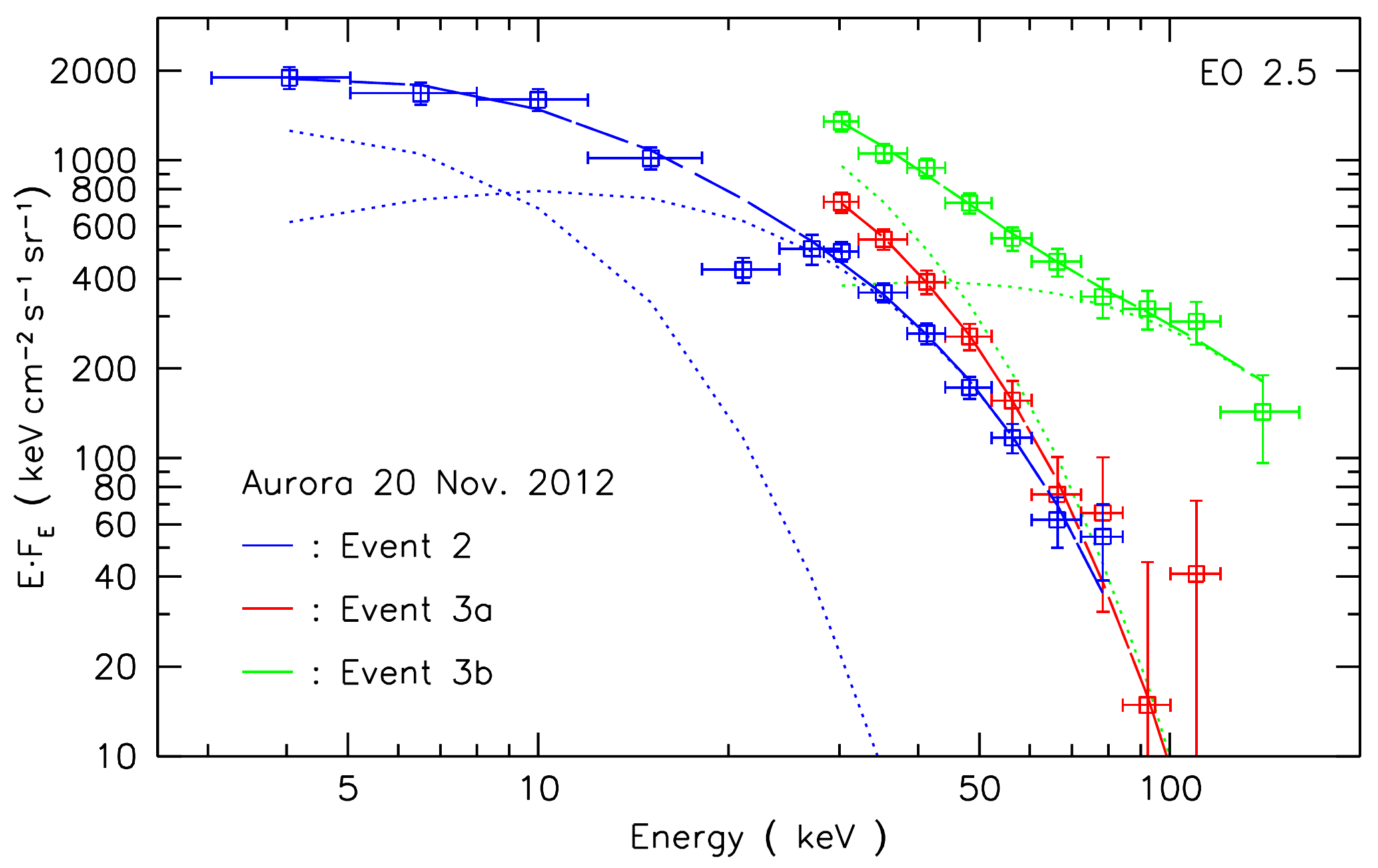}
\end{center}
\caption 
{ \label{fig:spectra2012}
Auroral spectra in IBIS/ISGRI and JEM-X on 20 November 2012. The spectra are labeled according to the time intervals they correspond to (see Fig.~\ref{fig:lightcurves2012}). JEM-X fluxes are the average of the fluxes obtained for JEM-X 1 \& 2, and are only available for Event 2. The fifth energy bin in JEM-X is apparently an outlier and was therefore disregarded for the fit. Dotted lines are used for the two distinct bremsstrahlung components of the fitted model (dashed line). All error bars include 5\,\% of systematic errors.
}
\end{figure} 

\subsection{Aurora on 20 November 2012}
\label{sec:aurorae_2012}
On 20 November 2012, \emph{INTEGRAL} observed between 14 and 22 hours (UT) a complex sequence of three distinct X-ray auroral events lasting typically between 1.5 and 2 hours (Fig.~\ref{fig:lightcurves2012}). The first one is apparently the brightest  in JEM-X, but radioactive decay in the spacecraft and residual contamination from the radiation belts at the start of the revolution are likely contributing to the strong count rates recorded before around 15 UT. We therefore limit the spectral analysis to the two other events.

A combined JEM-X and ISGRI spectrum is only possible for the second auroral event, which is rather soft such that useable ISGRI data are limited to energies below $\sim$80 keV (Fig.~\ref{fig:spectra2012}). A model composed of two bremsstrahlung components reproduces roughly the curved shape of the observed spectrum, but the fit has a much too high reduced $\chi^2$ ($\chi^2_{\mathrm{red}}$) of 7.5 for 15 degrees of freedom (d.o.f.). This is mainly due to the fifth energy bin of both JEM-X 1 \& 2 being outliers. Ignoring them lowers the $\chi^2_{\mathrm{red}}$ to 2.3 for 13 d.o.f., which can be brought down to 1.0 by further increasing the systematic uncertainties to 10\,\%. The first bremsstrahlung model then has a plasma temperature of $kT=4.9^{+1.1}_{-0.9}$\,keV and a normalisation of 1180$^{+220}_{-170}$ and the second $kT=15.7^{+1.4}_{-1.0}$\,keV with a normalisation of 530$\pm$100. Here and in the subsequent spectral analysis of aurorae, all indicated error ranges are 1\,$\sigma$ (68\,\% confidence).

The third auroral event starts with a prominent peak at low-energy in ISGRI. To study the suspected difference in spectral shape of this first part, we cut the event into two epochs `3a' and `3b'. The spectrum `3a' is indeed found to be very steep with a power law photon index of $\Gamma=4.43\pm0.13$ and a normalisation at 1\,keV of $(2.9\pm1.4)\times10^{6}$, whereas the subsequent spectrum `3b' is brighter with a harder photon index of  $\Gamma=3.33\pm0.06$ and a normalisation of $1.25^{+0.35}_{-0.27}\times10^{5}$. The simple power-law gives already a satisfactory fit for both spectra `3a' and `3b' with $\chi^2_{\mathrm{red}}$ of 1.14 for 7 d.o.f., and of 0.76 for 8 d.o.f., respectively. However, to get some idea about more physical quantities, a bremsstrahlung model can be used instead. For the spectrum `3a' a simple bremsstrahlung gives an excellent fit ($\chi^2_{\mathrm{red}}=0.36$) with a temperature of  $kT=13.8\pm 0.7$\,keV and a normalisation of $940\pm 110$. For spectrum `3b', one needs the addition of a second bremsstrahlung component to get an equally good fit as with the single power law ($\chi^2_{\mathrm{red}}=0.78$ for 6 d.o.f.). The two bremsstrahlung components then have temperatures of $kT=13.2\pm 2.6$\,keV and of $kT=64^{+31}_{-14}$\,keV with normalisations of $1320^{+610}_{-290}$ and of $108^{+55}_{-45}$, respectively.

%%% AURORA 2015 %%%

\begin{figure}[t]
\begin{center}
\includegraphics[width=1.0\linewidth]{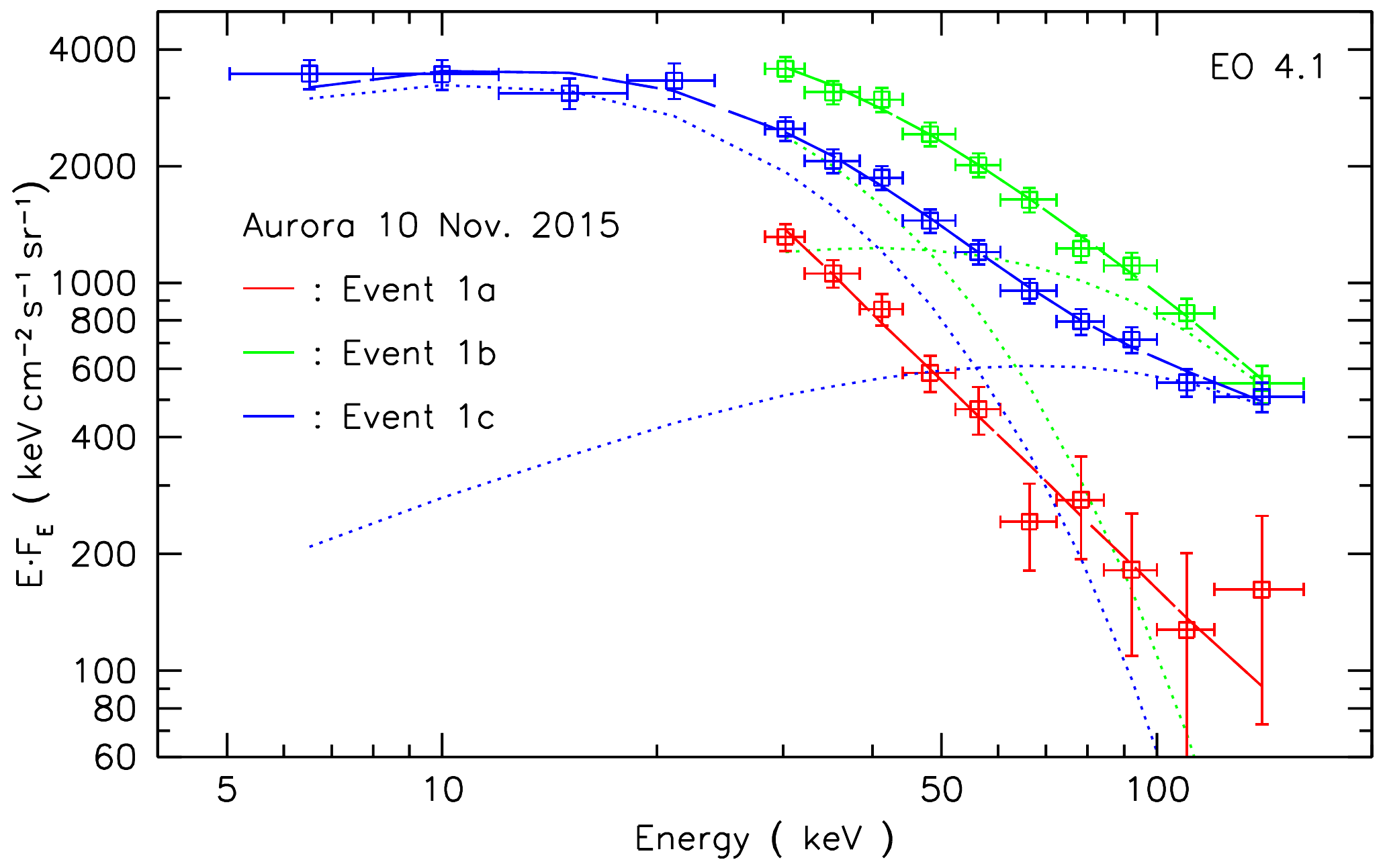}
\end{center}
\caption 
{ \label{fig:spectra2015}
Auroral spectra in IBIS/ISGRI and JEM-X on 10 November 2015. The spectra are labeled according to the time intervals they correspond to (see Fig.~\ref{fig:lightcurves2015}). As in Fig.~\ref{fig:spectra2012}, the fluxes of JEM-X 1 \& 2 have been averaged, error bars include 5\,\% of systematics, and dotted lines show the distinct bremsstrahlung components of the best-fit model (dashed line).
}
\end{figure} 

\subsection{Aurora on 10 November 2015}
\label{sec:aurorae_2015}
Another, even stronger, auroral event was observed by \emph{INTEGRAL} on 10 November 2015 between 10.5 and 15 hours (UT). The full event was recorded by IBIS/ISGRI, whereas the first part of the event was missed by JEM-X due to its smaller field of view. The light-curves in Fig.~\ref{fig:lightcurves2015} suggest a higher amplitude of variability on timescales of 10-15 minutes in JEM-X than in ISGRI. The image analysis suggests that the onset of the event comes from a distinct geographical location, alike for the event of 2012 (Fig.~\ref{fig:aurora_images}). We therefore define three time intervals for the spectral analysis.

Only the last part of the auroral event has reliable JEM-X data to be combined with ISGRI. We, however, had to disregard here the first and the last energy bins of JEM-X being apparent outliers. The combined fit of this spectrum `1c' has a reduced $\chi^2$ of 1.105 for 14 d.o.f. The first bremsstrahlung model has a plasma temperature of $kT=16.8\pm0.9$\,keV and a normalisation of 1730$\pm70$ and the second $kT=106^{+42}_{-11}$\,keV with a normalisation of 131$\pm$30.

The ISGRI data for the spectra `1a' and `1b' can be indicatively fitted with a single power-law to compare the spectral slopes. The spectrum `1a' corresponding to the onset of the event is fainter and has a softer photon index of $\Gamma=3.8\pm0.1$ with a normalisation at 1\,keV of $6.0^{+3.3}_{-2.1}\times10^{5}$, whereas the subsequent spectrum `1b' is brighter with a harder photon index of  $\Gamma=3.20\pm0.05$ and a normalisation of $2.3^{+0.5}_{-0.4}\times10^{5}$. The simple power-law gives already a satisfactory fit for the spectrum `1a' with $\chi^2_{\mathrm{red}}=0.85$. An alternative bremsstrahlung model gives here a less good fit ($\chi^2_{\mathrm{red}}=1.65$ for 8 d.o.f.), but indicates a temperature of $kT=18.4\pm 1.1$\,keV.  Spectrum `1b' on the other hand has some significant curvature such that the power-law model only gives a reduced $\chi^2$ of 2.14. The double bremsstrahlung model allows to properly fit the spectral curvature ($\chi^2_{\mathrm{red}}=0.68$), with plasma temperatures of $kT=18.4^{+4.5}_{-3.9}$\,keV and $kT=63^{+32}_{-12}$\,keV, and with normalisations of $1847^{+523}_{-282}$ and $344^{+186}_{-181}$, respectively.

\subsection{Discussion}
\label{sec:discussion}

Two strengths of \emph{INTEGRAL}, namely its long orbit enabling uninterrupted observations and its broad spectral energy range, were also essential here to study both the temporal and the spectral properties of auroral manifestations in the X-rays. Combined with imaging, it was possible to clearly identify on two occasions a short  spatially localised onset of the event on one side of the terrestrial globe with a rather soft spectrum. This first flare was then followed by a rapid move of the emission region to a wider area located eastward almost on the other side of the globe. This more extended emission lasted longer and is characterised by a harder spectrum that could be detected with \emph{INTEGRAL} up to 160\,keV.

The observed behaviour is typical for auroral substorms, which have properties already derived in the 1960s \citep{AH61,A64,B64}, see reviews by \citet{EMC96} and \citet{A17}. In the late 1990s the Polar Ionospheric X-ray Imaging Experiment (PIXIE) on-board the NASA Polar spacecraft was the first to be able to study simultaneously the entire auroral zone in X-rays. \citet{OSB99} review earlier balloon and low-altitude satellite X-ray measurements of substorms and confirm their typical development  with the unprecedented imaging capabilities of PIXIE in the $\sim$10--20\,keV range. It was found that a short substorm onset around midnight in Magnetic Local Time (MLT) is generally followed by an eastward expansion of the emission region to the morning sector with a maximum at 5--9 MLT and with a time delay of around half an hour consistent with drift models for electrons of $\sim$100\,keV. The statistical analysis of PIXIE observations by \citet{PCM99} shows, as expected, that the X-ray flux increases from low to high values of the \emph{Kp} index, which is a common measure of geomagnetic activity on a scale from 0 to 9. Moderately high \emph{Kp} values of 5 and just above 5 were measured, respectively, during the first and second  \emph{INTEGRAL} observations reported here\footnote{GFZ German Research Centre for Geosciences: \href{https://spaceweather.gfz-potsdam.de/products-data/nowcasts/nowcast-kp-index}{spaceweather.gfz-potsdam.de}}.

Spectral capabilities of PIXIE are quite limited both in energy range ($\sim$3--22 keV) and statistical uncertainties compared to \emph{INTEGRAL}. \citet{OSB01} show that a hard tail was often observed in PIXIE spectra, suggesting a double exponential electron spectrum. In one of the displayed PIXIE spectra the high-energy component is best fitted by bremsstrahlung from an electron spectrum with an $e$-folding high-energy cut-off energy at 80.4\,keV, suggesting that electron energies up to $\sim$100\,keV are rather common. Apart from PIXIE, nice images of terrestrial aurorae were obtained by the Chandra X-ray Observatory in the 0.1--10\,keV range in 2003--2004 \citep{BRE07}, but, at hard X-rays, the \emph{INTEGRAL} observations remain unprecedented so far. In a near future, the Atmospheric Effects of Precipitation through Energetic X-rays (AEPEX) CubeSat mission including a compact coded-mask instrument shall perform auroral spectral observations in the 50-300 keV range \citep{MXW20}.

%%% SOLAR FLARES %%%

\section{Solar Flares}
\label{sec:solar_flares}

From time to time, the Sun experiences huge surface explosions, which can convert in a few minutes up to $10^{32}$~erg of magnetic energy into accelerated particles, heated plasma, and ejected solar material \citep[e.g.,][]{rhessibook}. In the most intense solar flares (of class M or X), protons and heavier ions can be rapidly accelerated to kinetic energies of tens of GeV per nucleon, while electrons can reach at the same time hundreds of MeV. Some of these particles escape into the interplanetary space, and can pose a threat to satellite operations, while others are trapped in magnetic field loops at the surface of the Sun and produce a broadband non-thermal emission extending from radio waves to gamma rays by interacting with the solar atmosphere. Solar-flare gamma rays provide strong constraints on the acceleration mechanisms at work in the solar corona, as well as on the physical properties of solar active regions \citep{man00,vil11}. 

\subsection{High-resolution gamma-ray line spectroscopy}
\label{sec:spectroscopy}

\begin{figure}[ht]
\begin{center}
\begin{tabular}{c}
\begin{minipage}{0.5\columnwidth}
\includegraphics[scale=0.46]{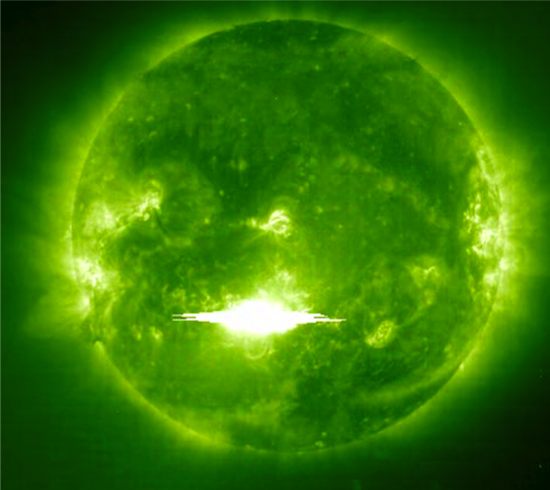}
\end{minipage}
\begin{minipage}{0.5\columnwidth}
\includegraphics[scale=0.29]{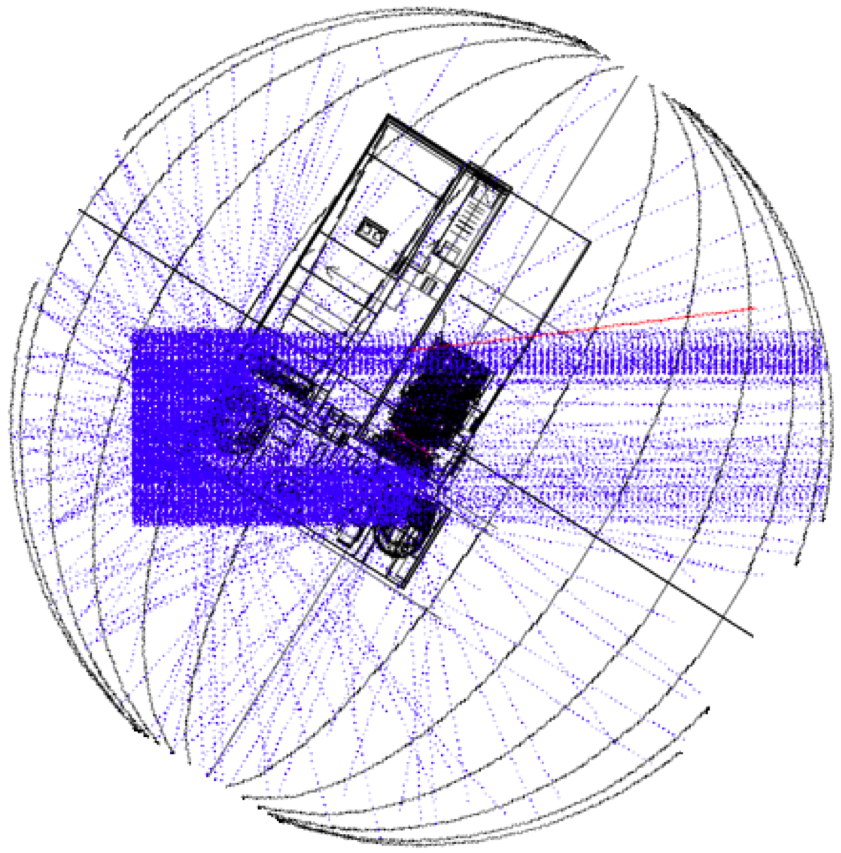}
\end{minipage}
\end{tabular}
\end{center}
\caption 
{ \label{fig:solar1}
({\it Left}) Image of the X17.2 class solar flare of October 28, 2003, observed with the Extreme Ultraviolet Imaging Telescope onboard the Solar and Heliospheric Observatory (SoHO). ({\it Right}) Monte-Carlo simulation of solar gamma-rays impinging the \textit{INTEGRAL} satellite during this flare, where the Sun was located at 122$^\circ$ from the satellite pointing direction. The {\it blue lines} are the photon tracks and the {\it red lines} the secondary particles created by photon interactions in the spacecraft (adapted from \cite{rosa14}).} 
\end{figure}

The Ge detectors of SPI can provide high-resolution solar-flare spectra in the 0.5--10~MeV range (solar photons below $\sim 500$~keV are absorbed in the satellite platform and/or IBIS, which is always located between SPI and the Sun to provide shadow for the SPI cooling system (see Fig.~\ref{fig:solar1}). The three most intense flares of the 23rd solar cycle have been studied with the SPI camera: SOL2003-10-28T11:10 (GOES class X17.2) \citep{gro04,kie06,wat06}, SOL2003-11-04T19:50 (X28.0) \citep{wat06,har07} and SOL2005-09-07T17:40 (X17.0) \citep{har07,wat07}. SPI spectra of two weaker flares of the current solar cycle were also extracted and analysed: SOL2012-03-07T00:24 (X5.4) and SOL2012-03-07T01:14 (X1.3) \citep{zha12}.  

\begin{figure}[t]
\begin{center}
\includegraphics[width=0.8\linewidth]{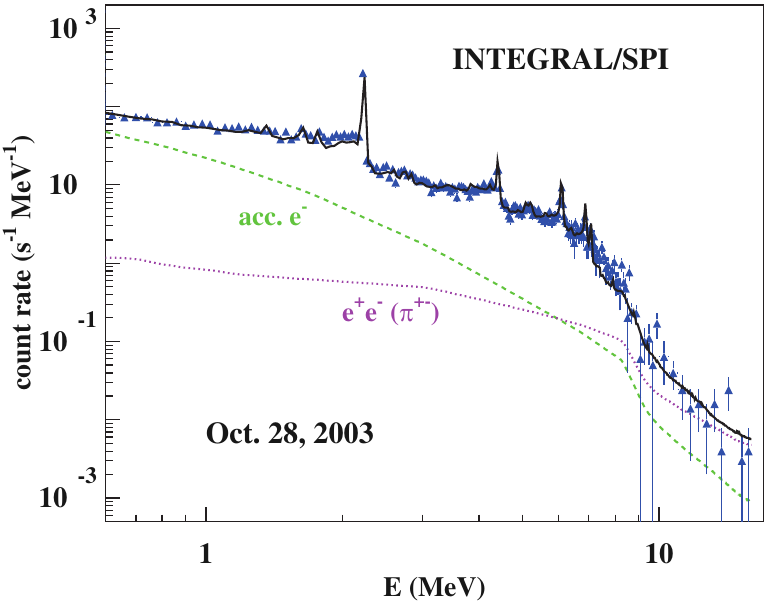}
\includegraphics[width=0.85\linewidth]{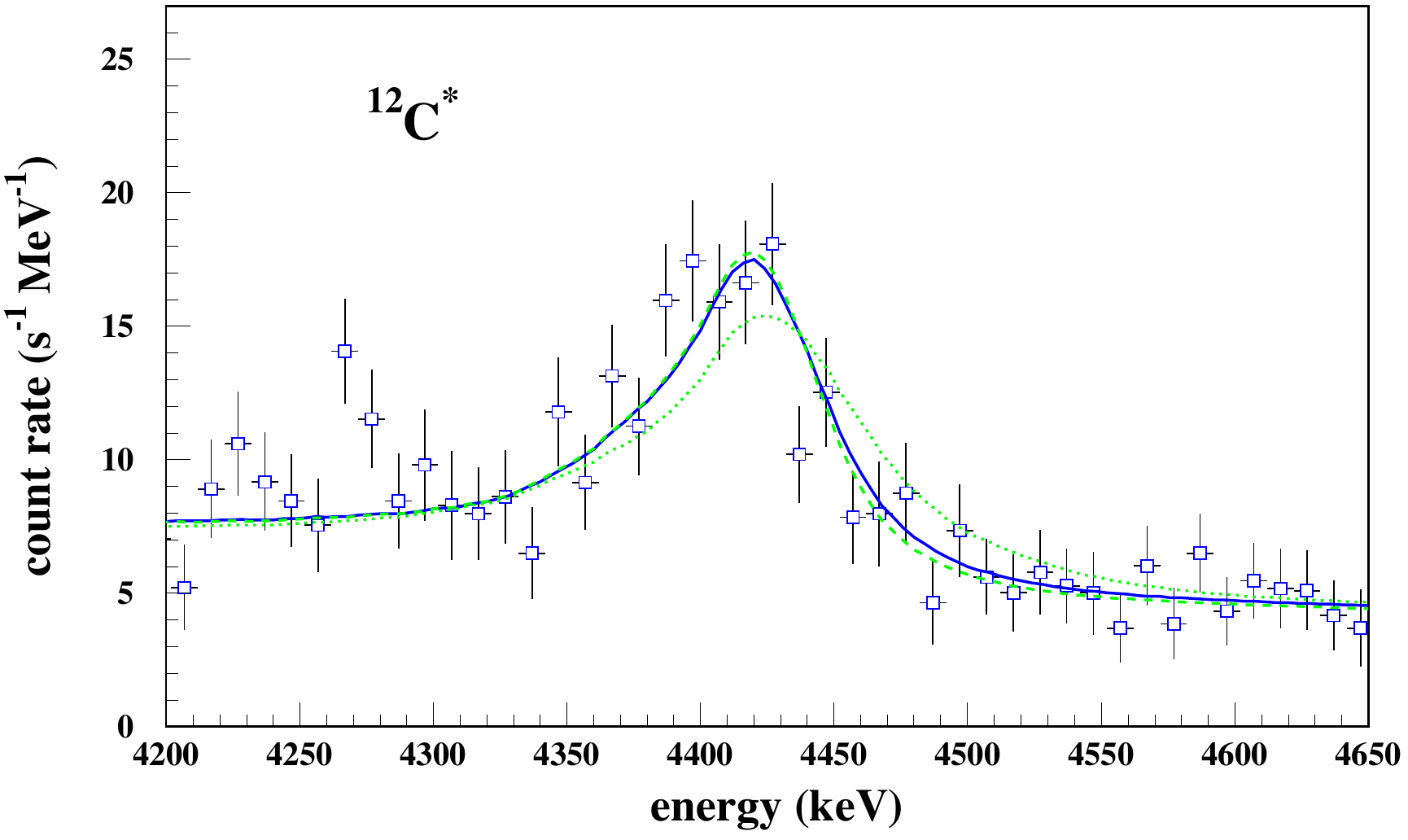}
\end{center}
\caption 
{ \label{fig:solar2}
({\it Top}) Gamma-ray spectrum of the October 28, 2003 solar flare as observed by \textit{INTEGRAL}/SPI during the most intense phase of the flare (from 11:02:13 to 11:14:13 UT). The spectrum is dominated by nuclear line emission ({\it solid line}). The modelled bremsstrahlung contributions of accelerated electrons and pion-decay leptons are shown by the {\it dashed} and {\it dotted lines}, respectively (adapted from \cite{kie12}). ({\it Bottom}) Zoom of the \textit{INTEGRAL}/SPI spectrum in the region of the 4.44~MeV ambient $^{12}$C deexcitation line. The {\it dashed} line represents the calculated line shape for a downward isotropic distribution of accelerated protons and $\alpha$-particles interacting in the solar chromosphere. The {\it solid} and {\it dotted lines} show calculated line shapes for a model of fast particle transport in the solar atmosphere with pitch-angle scattering on MHD turbulence \citep{mur07}, for two values of the normalised scattering mean-free path: $\lambda=30$ ({\it solid line}) and $\lambda=300$ ({\it dotted line}) (see \cite{kie06}).} 
\end{figure}

The strongest gamma-ray line in the SPI spectra of solar flares is generally the 2.223~MeV line from radiative capture of secondary neutrons by solar atmospheric hydrogen (see Fig.~\ref{fig:solar2}). The line is very narrow, 
%(intrinsic full width at half maximum $\lsim 1$~keV), 
because most of the secondary neutrons thermalise in the photosphere before being radiatively captured by protons almost at rest. The neutron capture line was found to be significantly weaker in the limb flares of November 4, 2003 (H$\alpha$ location S19W83) and September 7, 2005 (S06E89) \citep{har07}, 
%than in the other flares detected by SPI, 
because the flux of 2.223~MeV photons produced in the limb flares suffered more absorption in the solar atmosphere before reaching the \textit{INTEGRAL} spacecraft. However, high-energy neutrons of solar origin were detected by neutron monitors on Earth in association with all flares of solar cycle 23 detected by SPI \citep{wat06,wat07}. By using the gamma-ray emission time profiles measured with SPI as a proxy for the production rate of fast ions at the Sun, neutron energy distributions were deduced from the time profiles of the neutron-monitor count rates \citep{wat06,sak06,wat07,gon15}. These studies have shown that the detected neutrons have probably reached an energy of $\sim 1$~GeV, which implies that $\sim 10$~GeV protons were likely produced in these solar flares \citep{gon15}. Moreover, the flare of September 7, 2005 was found to produce high-energy neutrons over a more extended period than the flare impulsive phase detected by SPI/ACS in hard X-rays ($E \gsim 100$~keV; see below), which suggests that ions were continuously accelerated in some locations in the solar corona, or trapped longer in the emission site than the fast electrons \citep{sak06,wat07}.

The gamma-ray spectra of the strong solar flares detected by SPI also present two significant deexcitation lines produced by the interactions of flare-accelerated protons and $\alpha$-particles with the solar atmosphere: at 4.44 and 6.13~MeV from ambient $^{12}$C and $^{16}$O, respectively. The spectrum of the Oct. 28, 2003 flare (Fig.~\ref{fig:solar2}) also shows a line feature at $\sim 7$~MeV resulting from the decay of the $^{16}$O states at 6.92 and 7.12~MeV, as well as three weaker lines at 1.37, 1.63 and 1.78~MeV from ambient $^{24}$Mg, $^{20}$Ne and $^{28}$Si, respectively \citep{kie06}. All these lines are broadened by the recoil velocity of the excited nuclei, such that their shape depends on the composition and angular distribution of the fast particles in the interaction region, which in turn depends on the transport of these particles between their acceleration site, presumably located in the low corona, and the denser layers of the solar atmosphere (the chromosphere and photosphere), where the gamma-ray emission is produced. The transport of flare-accelerated ions is best described in a model where particles are injected isotropically into a coronal magnetic loop and propagate in the ambient magnetic field while undergoing pitch-angle scattering on magnetohydrodynamics (MHD) turbulence \citep[see][and references therein]{mur07}. 

Thanks to the high-resolution Ge detectors of SPI and the exceptional intensity of the X17 solar flare of Oct. 28, 2003, the SPI data for this flare are the best ever obtained to study the profile of solar gamma-ray lines. Detailed studies of the 4.44 and 6.13~MeV line shapes (see Fig.~\ref{fig:solar2} \textit{bottom}) have shown that the nuclear interactions were induced by downward-directed ion beams undergoing a high level of pitch-angle scattering \citep{kie06}. Thus, the SPI data have provided unique information on the level of MHD turbulence in solar-flare magnetic loops. Furthermore, from the line shapes, the accelerated alpha-to-proton ratio was measured to be of the order of 0.1, but the relative intensities of the $^{12}$C and $^{16}$O lines were found to vary during the flare, and this is best explained by a significant change in time of the $\alpha$/p ratio, rather than the abundances of the target species. 

\subsection{Unique determination of the solar photospheric $^3$He abundance}

\begin{figure}[t]
\begin{center}
\includegraphics[width=0.99\linewidth]{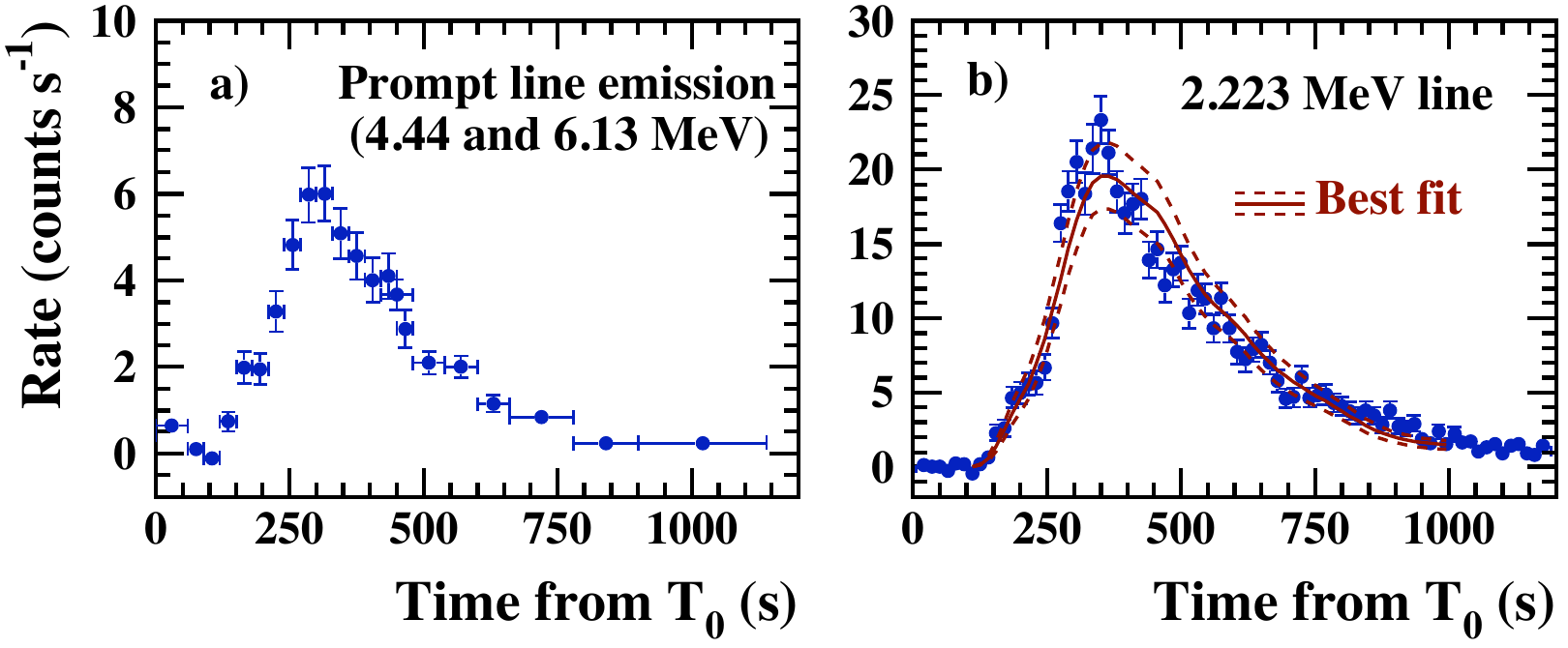}
\end{center}
\caption 
{ \label{fig:solar3} Light curves of ({\it a}) the sum of the 4.44 and 6.13~MeV lines and ({\it b}) the 2.22~MeV line measured by \textit{INTEGRAL}/SPI during the solar flare of October 28, 2003. Also shown in panel ({\it b}) is the best-fit model ($\chi^2_{\rm min}=61.5$ for 59 degrees of freedom) with $\pm 1\sigma$ uncertainties obtained for the photospheric abundance $(^3{\rm He/H})_{\rm ph}=1.5 \times 10^{-5}$ (see text). The reference time of the light curves is $T_0 = 11$h00m~UT on Oct. 28, 2003.
} 
\end{figure}

The knowledge we gained on the conditions of ion acceleration and transport in 
the flare of Oct. 28, 2003 makes this event unique for an estimate of the abundance of photospheric $^3$He from the measured time development of the 2.223~MeV line (Fig.~\ref{fig:solar3}b). Time history measurements of this line can provide a unique determination of the ambient $^3$He/H ratio (the He photospheric abundance cannot be measured by atomic spectroscopy), because non-radiative capture of neutrons by $^3$He via the $^3$He($n$,$p$)$^3$H reaction can significantly shorten the delay of the radiative-capture line emission via the $^1$H($n$,$\gamma_{2.223}$)$^2$H reaction \citep{wan74}. Here, we update the calculations first presented in \cite{tat05} by including all the information available from the gamma-ray line shape analysis \citep{kie06} and the neutron-capture line imaging of \textit{RHESSI} \citep{hur06}. The image at 2.22~MeV observed with \textit{RHESSI} during this flare shows two compact sources separated by $\sim 75''$ ($D\approx 55\,000$~km at the distance of the Sun), which most likely correspond to the two foot-points of the flare magnetic loop. Assuming that the magnetic loop has a semicircular shape in its coronal portion, the loop length is $L = \pi D/2 \approx 85\,000$~km. 

Adopting the measured time profile of the prompt gamma-ray line flux (Fig.~\ref{fig:solar3}a) as a proxy for the time history of neutron production at the Sun, we used the Monte-Carlo code developed by \citet[][and references therein]{hua02} to model the neutron transport and capture in the solar atmosphere. We assumed that the ambient medium where the neutrons are produced has a coronal composition and that the fast ion population has the composition of solar energetic particles measured in impulsive flare events from interplanetary space \citep{rea99}. With the fast ion transport parameters derived from the gamma-ray line shapes \citep{kie06}, the only free parameter of the model is then the $^3$He photospheric abundance. The best-fit model to the measured light curve of the neutron capture line is obtained for $(^3{\rm He/H})_{\rm ph}=(1.5_{-1.0}^{+1.6}) \times 10^{-5}$, where the errors are at the 90\% confidence level and take into account the statistical uncertainty on the prompt gamma-ray line flux (Fig.~\ref{fig:solar3}a). This result is in good agreement with the protosolar abundance $^3{\rm He/H}=(1.61 \pm 0.09) \times 10^{-5}$ that can be estimated from the He isotopic ratio measured by the \textit{Galileo} probe in the Jovian atmosphere, $^3{\rm He/^4He}=(1.66 \pm 0.05) \times 10^{-4}$ \citep{mah98}, and the $^4$He protosolar abundance $^4{\rm He/H}=0.0969 \pm 0.0046$ \citep{lod09}. However, the present-day abundance of $^3$He in the photosphere might be different from the protosolar value because of two effects: (\textit{i}) deuterium burning in the solar core via the D($p$,$\gamma$)$^3$He reaction, which might have increased the $^3$He abundance in the solar atmosphere and (\textit{ii}) mass-dependent settling from the photosphere into the Sun's interior, which might have decreased the $(^3{\rm He/H})_{\rm ph}$ ratio. The fact that the $^3$He photospheric abundance measured with \emph{INTEGRAL} is compatible with the protosolar abundance suggests that these two effects did not play an important role for the current-day composition of the photosphere or that they compensated for each other since the birth of the Sun 4.6 billion years ago.

\subsection{SPI/ACS and IREM for space weather}

The Anti-Coincidence Shield (ACS) of the SPI instrument also provides valuable data for solar flare physics. The response of SPI/ACS for several selected flares detected in nonstandard observation conditions has been carefully studied through Monte Carlo simulations \citep{rosa14}. Within the European Union's FP7 SEPServer project \citep{vai13}, ACS light-curves have been extracted and studied for a set of 24 solar flares which occurred between 2002 and 2006 \citep{rosa14}. In the frame of the EU's H2020 HESPERIA project \citep{hesperia}, 25 more solar flare light curves measured with SPI/ACS between 2011 and 2014 were provided to the solar physicist community \citep{kle18}. These two European projects were dedicated to the study of Solar Energetic Particle (SEP) events with the aim of providing a public database of solar events and forecasting tools for space weather, using a large set of SEP data and associated electromagnetic emissions  at the Sun from radio to gamma-ray frequencies. 

Thanks to its large effective area \citep[$A_{\rm eff}\sim 100$~--~$1000$~cm$^2$ depending on the flare observation conditions; ][]{rosa14}, SPI/ACS can provide high-quality hard X-ray/gamma-ray light curves for X- and M-class solar flares and, in some case, probably also for C-class flares. Detailed comparisons of  ACS data with time evolutions of soft X-ray and radio emissions have shed a new light on the generation of highly energetic electrons in the solar corona and the transport of these particles to the chromosphere, where they emit hard X-ray and gamma-ray radiation via bremsstrahlung \citep[e.g.,][]{man09,zim12,str13}. During the impulsive and early post-impulsive phase of solar flares, high-energy protons and $\alpha$-particles $>$$300$~MeV/nucleon can be accelerated concomitantly with electrons and generate a characteristic hump at photon energies above 60 MeV by pion production and decay. But the late-phase of $>$$100$~MeV gamma-ray emission observed by the \textit{Fermi} Large Area Telescope (LAT) lasts much longer than the impulsive hard X-ray emission phase traced by SPI/ACS  \citep{share18,kle18}, and the origin of these long-duration gamma-ray events is still debated. Associated to data from the Helioseismic and Magnetic Imager (HMI) on board the \textit{Solar Dynamics Observatory}, the ACS data are also used to study the generation of helioseismic waves (also referred to as ``sunquakes'') during strong solar flares \citep{sha18}. 

Lastly, the \textit{INTEGRAL} Radiation Environment Monitor (IREM) provides useful information to understand the solar origin, acceleration, and propagation of SEP events. IREM belongs to the second generation of ESA's Standard Radiation Environment Monitor (SREM), and similar units have also been operational onboard the \textit{STRV-1C}, \textit{Proba-1}, \textit{Rosetta}, \textit{GIOVE-B}, \textit{Herschel} and \textit{Planck} satellites. It was shown to be a reliable particle radiation monitor (measuring electrons above 500~keV and protons above 10 MeV) able to detect major solar particle events associated with large M- and X-class flares and possible accompanying coronal mass ejections \citep{tzi10,geo18}. Associated to other SREM units onboard spacecrafts at different locations, it could be part of an efficient alert system for explosive geoeffective solar events \citep{pap11}, possibly including the prediction of magnetospheric disturbances \citep{geo18}. 

\section{Radiation belts}
\label{sec:radiation_belts}

The environment monitor IREM \citep{HBE03} and SPI's active shielding (ACS) can be used to gain insight into the extent, energetics and variability of the Earth's radiation belts. 
Thanks to \textit{INTEGRAL}'s highly eccentric orbit, the radiation belts are crossed about every 3 days, while it remains outside the radiation belts for about 90\% of its orbit to reduce the background on its main scientific instruments.  \citet{HBE03} demonstrated the capabilities of IREM to measure for example the location of the electron belt entry and exit regions, as well as particle rates and rates ratio as a function of solar events. 

\citet{MHS17} used 14~years of IREM's data to study the electron fluxes in the 0.5--2\,MeV range measured at different locations in the radiation belts and to correlate these data with other space environment values, such as the sunspot number, solar wind velocity and solar wind pressure. 
They performed an extreme value analysis and found that fluxes of highly relativistic electrons during extreme high flux periods are likely to be a factor of  3--5 times higher near the equator in medium Earth orbit than at geosynchronous orbit, depending on energy. 
This goes along with spectral hardening of the electron energy distribution when the relativistic flux rises.  Studies like the one of \citet{MHS17} show that \textit{INTEGRAL} provides a rich data set that allows space environment studies far beyond the original scope of this soft gamma-ray mission.

\section{Conclusion}
\label{sec:conclusion}

The serendipitous science of \textit{INTEGRAL} includes sources in the solar system, which were not identified as potential science targets before the launch.  By relaxing the constraint to always point away from the Earth, \textit{INTEGRAL} had the chance to record on two occasions relatively bright X-ray auroral activity, which are presented here for the first time and provide a set of unique hard X-ray spectra of auroral substorms. \textit{INTEGRAL} was never allowed to directly point towards the Sun, but nevertheless had a glimpse from the side at bright solar flares. Thanks to the high resolution of SPI's Ge detectors, some of the SPI data are the best ever obtained to study in detail the profile of solar gamma-ray lines. Finally, the very long life of the spacecraft and its changing orbit over time allowed to study the radiation belts, thus turning this nuisance for spacecraft operations into a new science target.


 \section*{References}

  \bibliographystyle{elsarticle-harv} 
  \bibliography{serendipitous}


%\begin{thebibliography}{00}



%\end{thebibliography}

\end{document}